\documentclass[aps,prb,twocolumn,superscriptaddress,fontsize=10pt]{revtex4-2}

\usepackage{bm}
\usepackage{amssymb}
\usepackage{amsmath}
\usepackage{siunitx}
\usepackage{textcomp}
\usepackage{graphicx}
\usepackage{natbib}
\usepackage{color}
\usepackage{setspace}
\usepackage{comment}

\newcommand{\sld}[1]{\mathrm{#1}}

\begin{document}

\title{Internal structure of localized quantized vortex tangles}


\author{Tomo Nakagawa}
\author{Sosuke Inui}
\affiliation{Department of Physics, Osaka City University, 3-3-138 Sugimoto, 558-8585 Osaka, Japan}
\author{Makoto Tsubota}
\affiliation{Department of Physics, Osaka City University, 3-3-138 Sugimoto, 558-8585 Osaka, Japan}
\affiliation{Department of Physics, Nambu Yoichiro Institute of Theoretical and Experimental Physics, and OCU Advanced Research Institute for
Natural Science and Technology, Osaka City University, 3-3-138 Sugimoto, Osaka 558-8585, Japan}


\date{\today}

\begin{abstract}
In this study, we numerically investigate the internal structure of localized quantum turbulence in superfluid $^4$He at zero temperature with the expectation of self-similarity in the real space. In our previous study, we collected the statistics of vortex rings emitted from a localized vortex tangle. As a result, the power law between the minimum size of detectable vortex rings and the emission frequency is obtained, which suggests that the vortex tangle has self-similarity in the real space [Nakagawa $et$ $al.$, Phys. Rev. B $\bm{101}$, 184515 (2020)]. In this work, we study the fractal dimension and vortex length distribution of localized vortex tangles, which can show their self-similar structure. We generate statistically steady and localized vortex tangles by injecting vortex rings with a fixed size. We used two types of injection methods that produce anisotropic or isotropic tangles.  The injected vortex rings develop into a localized vortex tangle consisting of vortex rings of various sizes through reconnections (fusions and splitting of vortices).
The fractal dimension is an increasing function of the vortex line density and becomes saturated to a value of approximately 1.8, as the density increases sufficiently. The behavior of the fractal dimension was  independent of the anisotropy of the vortex tangles. The vortex length distribution indicates the number of vortex rings of each size that are distributed in a tangle. The distribution of the anisotropic vortex tangle shows the power law in the range above the injected vortex size, although the distribution of the isotropic vortex does not. 
\end{abstract}


\maketitle

\section{Introduction}
Turbulence is one of the most important problems in physics, and many  researchers have studied it for many years \cite{Frisch,Davidson}.  
It is a typical phenomenon that is non-equilibrium and non-linear, which prevents us from understanding it well. Quantum turbulence is noticeable to resolve this problem.
In classical turbulence or conventional fluid turbulence, it is difficult to identify vortices, which are important elements of turbulence, because they are unstable and not well defined. However, in quantum turbulence, namely turbulence in quantum condensed fluids, vortices are well-defined topological defects and have quantized circulation. Therefore, studies on quantum turbulence provide a shortcut to understanding turbulence \cite{Vinen2002,Tsubota2013,Barenghi2014,Tsubota2017}.

Superfluid $^4$He is a typical quantum fluid. Liquid $^4$He transitions to the superfluid phase below the temperature $T_\lambda=2.17\ \mathrm{K}$. The physics of superfluid $^4$He is described by the two-fluid model; the system consists of a superfluid component and a normal fluid component \cite{Tisza1938,Landau1941}. The superfluid component has no viscosity and entropy, whereas the normal fluid component has them. The ratio of the densities depended on the temperature. 
 When the temperature was below $1\ \mathrm{K}$, the normal fluid component almost disappeared. 
 In the superfluid component, the circulation is quantized to $\kappa=h/m$, where $h$ and $m$ are the Planck constant and the mass of a $^4$He atom, respectively \cite{Vinen1961}. A typical quantum turbulence consists of a tangle of quantized vortices.

In turbulence, the velocity field is disturbed both spatially and temporally.
Therefore, researchers have tried to describe turbulence not by the velocity field but by universal and reproducible statistical laws.
In classical turbulence, Kolmogorov's law is well known as the universal law showing self-similarity in the wave number space. This law implies that the energy spectrum of isotropic and homogeneous turbulence is proportional to the wave number to the power of $-5/3$, and its energy is transferred from low wave numbers to high wave numbers. Kolmogorov's law has also been studied and confirmed in quantum turbulence \cite{NorePRL,NorePoF,Maurer1998,Stalp1999,Kobayashi2005,Kobayashi20052}. 
However, the self-similarity in the real space is not well understood because the vortices are not well defined in classical turbulence, although several self-similar models have been studied, such as the $\beta$-model \cite{Frisch}. In quantum turbulence, all vortices are well-defined, which enables us to consider the self-similarity in the real space clearly.

The quantum turbulence of superfluid $^4$He has been studied for more than half a century. 
Most efforts of researches have been devoted to thermal counter flow \cite{Vinen2002,Tsubota2013}.
However, in recent years, localized turbulence, such as that generated by oscillating objects, has received considerable  attention \cite{Jager1995,Luzyriaga1997,Yano2007,Hashimoto2007,Garg2012,Bradley2009,Bradley20092,Yano2010,Bradley2011,Bradley2012,Bradley20123,Yano2019}. A typical example of an oscillating object is a vibrating wire \cite{Yano2007,Yano2010,Yano2019}. Yano $et\ al.$ conducted wire experiments and found some statistical laws for localized vortex tangles. They generated the vortex tangle by the vibrating wires and obtained the statistics of vortex rings emitted from the tangle.  Surprisingly, they obtained laws that may represent self-similarity in real space. They found a power law between the minimum size of
detectable vortex rings and their detection frequencies. It may show self-similarity in the real space because the emitted vortex rings provide information on the structure of the tangle where they belong.

Our goal is to understand self-similarity in the real space of localized vortex tangles.  
In our previous study \cite{Nakagawa2020}, we investigated vortex rings emitted from vortex tangles and obtained results qualitatively similar to the experiments in a previous study \cite{Yano2019}, such as the self-similarity of the  vortex emission . 

In this study, we develop the previous study and investigated the internal structure of vortex tangles with the expectation of self-similarity. 
To accomplish this goal, we studied the fractal dimension \cite{Vassilicos1996,Kivotides2001}  and the vortex length distribution \cite{Araki2002,Fujiyama20102},  which can show the self-similarity of vortex tangles directly.
The fractal dimension is a non-integer dimension that characterizes a self-similar structure. This dimension was calculated by Kivotides $et$ $al.$ for vortex tangles in periodic boundary conditions \cite{Kivotides2001}. The vortex length distribution is a probability density function of the vortex rings in a vortex tangle depending on their size. The distribution is expected to represent a self-similar turbulence structure. In Sections III and IV, we describe them in detail.

We also investigated the universality independent of the anisotropy of the tangles. We made the tangles using two methods numerically. One is the same method as in our previous study, which makes anisotropic localized vortex tangles \cite{Nakagawa2020}.  The other is to create isotropic localized vortex tangles. The two methods enable us to investigate the universality independent of the anisotropy of the tangle.

This paper comprises five sections. In Section II, we introduce the vortex filament model and the system that we consider. Section III presents the results of calculating the fractal dimensions. In Section IV, we discuss the vortex length distribution. Finally, the conclusions are presented in Section V.

\section{The model and system}
\subsection{Vortex filament model}
The circulation of vortices in superfluid $^4$He is quantized, and the cores have very thin structures of the order of $ 1\ \AA$. 
For these reasons, the thin structures of the vortex cores can be neglected, and a vortex filament model is available \cite{Schwarz1985}. In this study, we performed a simulation of the dynamics of the vortex filament model at $T=0\ \mathrm{K}$. The superfluid velocity field obtained by the filaments obeys the Biot-Savart law. The equation of motion of the vortex filaments is:
\begin{equation}\label{eq:BS}
\frac{d\bm{s}}{dt}=\frac{\kappa}{4\pi}\int_L\frac{\bm{s}^\prime(\xi,t)\times(\bm{r}-\bm{s}(\xi,t))}{\mid\bm{r}-\bm{s}(\xi,t)\mid^3}d\xi \;,
\end{equation}
where $\bm{s}(\xi,t)$ is the position of the filaments represented by the parameter $\xi$, and $\bm{s}^\prime$ is $\partial \bm{s}/\partial \xi$. The integration was performed over all vortex filaments $L$. 

To calculate this equation numerically, we discretized the vortex filaments into points. The eq. (\ref{eq:BS}) is solved by fourth order Runge-Kutta method. As this model cannot describe the reconnection of vortices, we make them reconnect algorithmically when their distance is less than  $\Delta \xi$=$0.5\ \mathrm{\mu m}$. We deleted small vortex rings with sizes smaller than $5\Delta \xi$ as dissipation.

\subsection{The systems}

We generate localized vortex tangles by two different methods, ``parallel injection" and ``spherical injection". The advantage of these methods is that they can locally generate a steady vortex tangle. The steady state of the vortex tangle indicates the equilibrium between the excitation and dissipation in a finite volume because we are interested in a localized vortex tangle. The method of determining the volume occupied by the tangle depends on the injection methods and will be introduced later.  The excitation originates from vortex injections. There are two types of dissipation: the vortex ring escaping from the finite volume and the deletion of the small vortex rings. The ratios of these two types of dissipation were roughly the same.
\subsubsection{parallel injection}

\begin{figure}[h]
\includegraphics[scale=0.2]{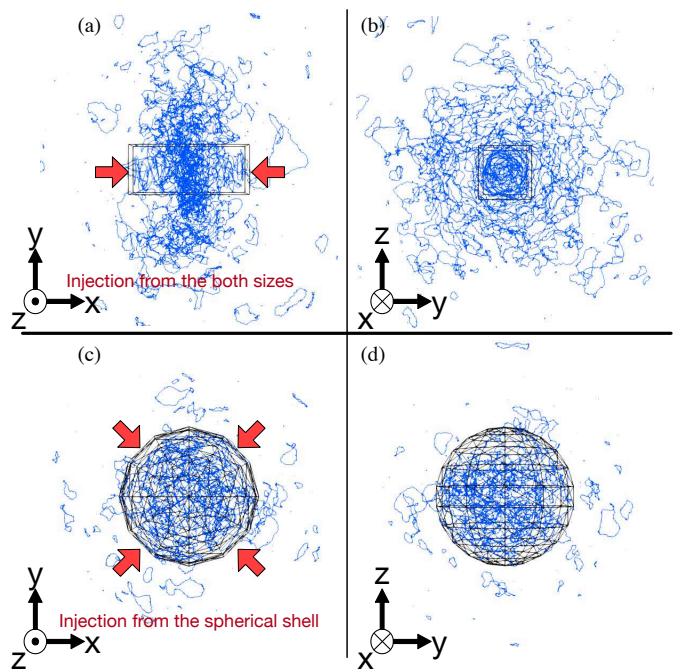}
\caption{\label{fig:vortextangle}The vortex tangles generated by the vortex injection at time $0.2\ \mathrm{s}$. (a) and (b) are the vortex tangle generated by ``parallel injection" in the condition $2R_0=60\ \mathrm{\mu m}$ and $f=1000\ \mathrm{Hz}$.  Moreover, (c) and (d) are the  tangle generated by ``spherical injection" in the condition $2R_0=60\ \mathrm{\mu m}$ and $f=1000\ \mathrm{Hz}$. }
\end{figure}
The ``parallel injection" method was used in our previous study \cite{Nakagawa2020}. In this model, two parallel square vortex sources are prepared, and vortex rings are injected from a random position of the two sources simultaneously with a certain frequency $f$ to collide them and generate a localized tangle.  The parameters of this model are the njection frequency $f$ and diameter $2R_0$ of the injected rings. The vortex tangles made by this method are anisotropic, as shown in Fig. \ref{fig:vortextangle}(a) and (b). In this system, the vortex line length in a finite volume covering the tangle becomes statically steady after a certain time, as shown in Fig. \ref{fig:ld}(a). The finite volume we applied is a cylindrical volume with a height of $160\ \mathrm{\mu m}$ and radius $250\ \mathrm{\mu m}$.  As we are interested only in the localized tangle, we deleted the vortex rings that escape far from the tangle. 

\subsubsection{spherical injection}

\begin{figure}
	\includegraphics[scale=0.25]{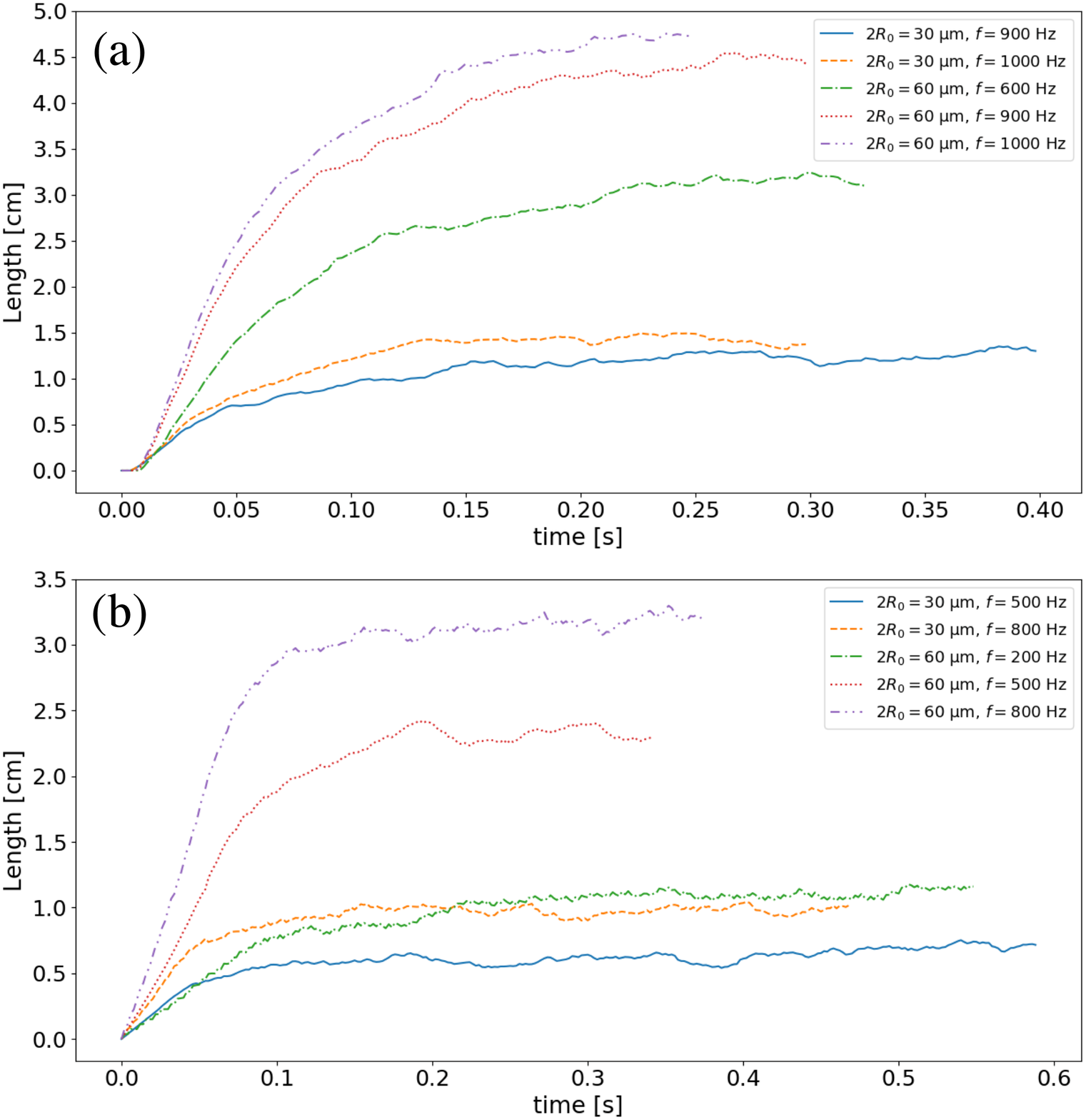}
	\caption{\label{fig:ld}Time development of the vortex line length. (a) is the tangle generated by ``parallel injection" and (b) is generated by ``spherical injection". The length is calculated in the finite volume. In the case of ``parallel injection", the volume is the cylindrical volume with height $160\ \mathrm{\mu m}$ and radius $250\ \mathrm{\mu m}$. The volume in ``spherical injection" is the spherical volume with radius $180\ \mathrm{\mu m}$. Both volumes cover the vortex tangles.}
\end{figure}

In the ``spherical injection" method, we inject vortex rings from the virtual spherical source of radius $180\ \mathrm{\mu m}$. An example of the tangle is shown in Fig. \ref{fig:vortextangle}(c) and (d). The parameters are similar to those of the former method. With a fixed frequency $f$, two vortex rings with size $2R_0$ are injected from random positions on the source. The tangles in a finite volume become statistically steady after a certain time, as shown in FIg. \ref{fig:ld}(b). In this method, the applied finite volume is a spherical volume with radius $180\ \mathrm{\mu m}$.  We also deleted vortex rings that are far from the tangle. 

\begin{figure}[h]\hspace{-27pt}
	\includegraphics[scale=0.28]{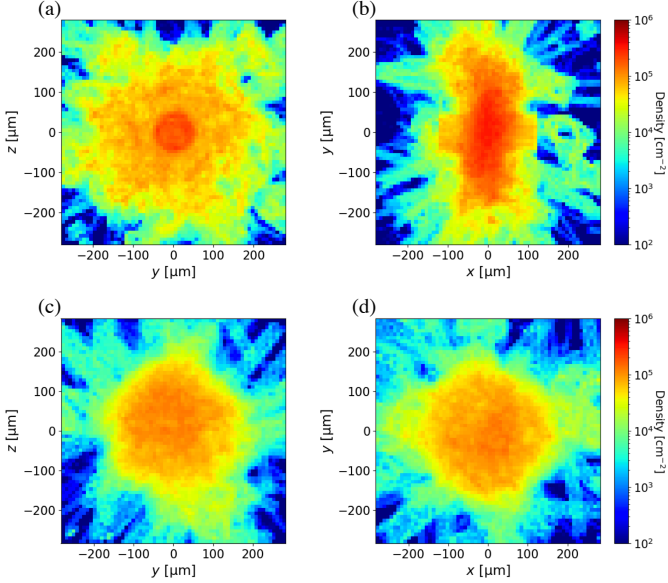}
	\caption{\label{fig:densdist}Time averaged vortex line density distribution of the vortex tangles. (a)-(b) are the distribution of ``parallel injection", while (c)-(d) are the distribution of ``spherical injection''.  The density is calculated in $280\mathrm{\mu m}\times\Delta y\times\Delta z $ in (a) and (c). (b) and (d) are the density in $\Delta x \times \Delta y \times 280\mathrm{\mu m}$. Here, $\Delta x =\Delta y=\Delta z =280/64 \mathrm{\mu m}$}
\end{figure}

The vortex line density distributions of the tangles are shown in Fig. \ref{fig:densdist}. The tangle generated by ``parallel injection" spreads out in the $y$ and $z$ directions and is relatively flat along the $x$-axis. However, that by ``spherical injection" is isotropic. The purpose of using the two methods is to determine the universality independent of the anisotropy.

\section{fractal dimension}
A fractal dimension is a non-integer dimension that represents how the fine structure of a pattern changes with the measured scale defined by 
\begin{equation}\label{eq:fracdim}
N\propto k^{D_f},
\end{equation}
where $k$ is a scale and $N$ is the number of structures with scale $k$.
We obtain the fractal dimension of our vortex tangles using the box-counting method \cite{Vassilicos1996,Kivotides2001}.
In this method, we divide a virtual box with the size of the system into a scale of length $\delta$ and count the number $N(\delta)$  of the boxes covering the tangle. Therefore, the fractal structure should show the following relation:
\begin{equation}\label{eq:box}
D_f=-\frac{\ln{(N(\delta)/N_\sld{min})}}{\ln{(\delta/\delta_\sld{min})}},
\end{equation}
where $\delta_\sld{min}$ is the minimum scale of length of the fitting range, and $N_\sld{min}$ is $N(\delta_\sld{min})$.
By plotting $\delta$ and $N(\delta)$ in a log-log scale and calculating its slope, we can determine the fractal dimension of the pattern. In this calculation, the maximum size of the box is set to $800\ \mathrm{\mu m}$ per side.

Regarding the fractal dimension of quantum turbulence, there is a previous study \cite{Kivotides2001}. Kivotides $et$ $al.$ prescribed a normal fluid turbulent flow and generated vortex tangles numerically through mutual friction under periodic conditions. Then, they obtained the fractal dimension of the vortex tangles with a density of $0.05$ - $0.20\times10^5\ \mathrm{ cm}^{-2}$ by the box-counting method and  discovered that the dimension $D_f$ is an increasing function of the vortex line density.  Their values of $D_f$ were roughly 1.4 - 1.7.

\begin{figure}[h]
\includegraphics[scale=0.5]{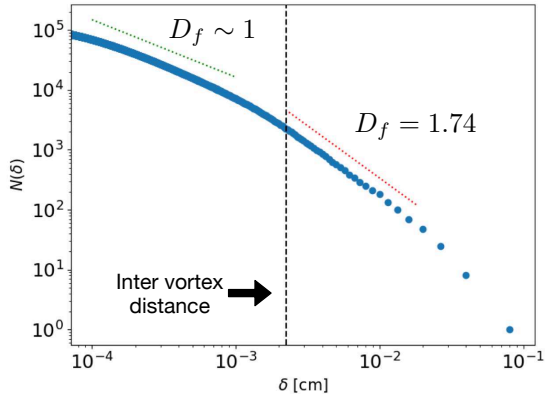}
\caption{\label{fig:60_1000count}The plotting of $\delta$ and $N(\delta)$ applying the box-counting method to the tangle by ``parallel method" in the condition $2R_0=60\ \mathrm{\mu m}$ and $f=1000\ \mathrm{Hz}$ ($\delta_\mathrm{min}=2.22\times 10^{-3}\ \mathrm{cm}$ and $N_\mathrm{min}=2526$). The green line shows the slope in the small scale(from the inter vortex distance to the space resolution). The red one shows the slope in the middle scale(up to the inter vortex distance).}
\end{figure}

In this study, we calculate the fractal dimension of localized vortex tangles using the box-counting method. 
Figure \ref{fig:60_1000count} shows the typical result in a statistical steady state of ``parallel injection". In small scales, it shows $D_f\approx 1 $, which refers to the dimension of each vortex filament as one-dimensional object. 
When the scale exceeds a certain critical value, $D_f$ changes to another value larger than unity, which is the fractal dimension of the tangle.
This behavior is confirmed for all vortex tangles produced under different conditions. We define the value at the large scale as the fractal dimension $D_f$ of the vortex tangle.  

The transition from a small scale to a large scale is determined by the mean intervortex distance.
The distance is the square root of the inverse of the vortex line density. The density is calculated with the vortex line length in a finite volume enclosing a central region of the tangle, as mentioned in Section II(B).

\begin{figure}[h]
\includegraphics[scale=0.5]{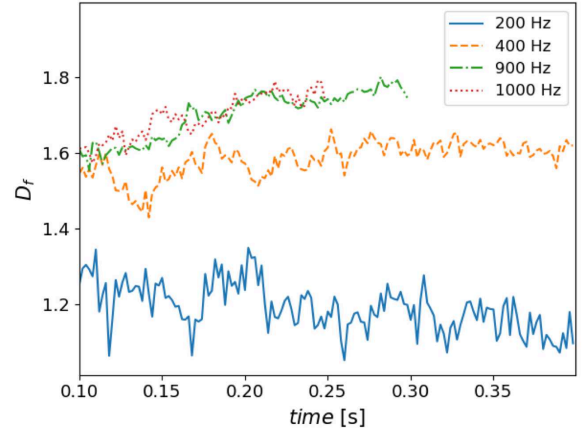}
\caption{\label{fig:jikanhatten}The time-development of the fractal dimension of the vortex tangles generated by ``parallel injection" in the conditions $2R_0=60\ \mathrm{\mu m}$ and some injection frequencies $f$. }
\end{figure}
Figure \ref{fig:jikanhatten} shows the time-development of $D_f$ for ``parallel injection". In earlier times, they fluctuate because the systems have a small number of vortices that it is difficult to calculate the fractal dimension of vortex tangles properly. 
\begin{figure}[h]
\includegraphics[scale=0.50]{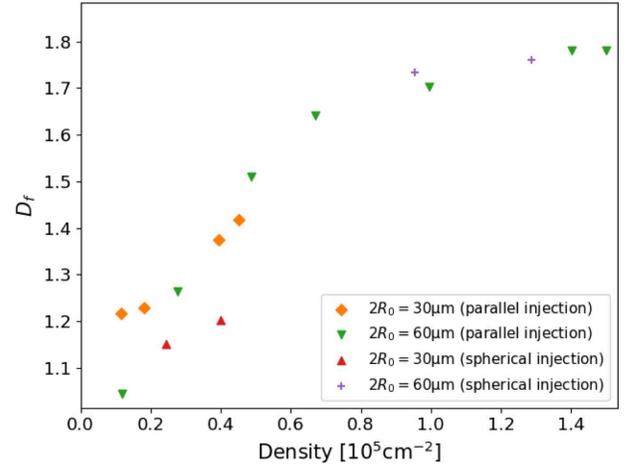}
\caption{\label{fig:souzu}The relation between the vortex line density and the fractal dimension. The dimension becomes saturated as density increases.}
\end{figure}

However, at a later time, $D_f$ becomes stationary and asymptotically approaches a certain value in each condition, which indicates that the tangles become statistically steady. Figure \ref{fig:souzu} shows the time-averaged $D_f$ after 0.2 s as a function of vortex line density.
In the low-density range, $D_f$ increases with density. This behavior is consistent with the results of Kivotides $et$ $al.$ \cite{Kivotides2001}.
In the higher density range, the fractal dimensions converge to $D_f\sim 1.8$, regardless of the injection method. Although the tangles have completely different shapes depending on their injection method, the tangles have similar fractal dimensions. This is non-trivial. The developed vortex tangles construct  intermittent structures on their own owing to the interaction between vortices, and the tangles show similar fractal dimensions $D_f\sim 1.8$. The values of the fractal dimension roughly agree with the results of Kivotides $et$ $al.$ \cite{Kivotides2001}. However, it is difficult to make a quantitative comparison because the systems are different. 

\section{vortex length distribution}
In this calculation, all vortices are closed loops because the system is free from solid boundaries. 
The vortex length distribution is the probability density function of the existence of vortex rings in the tangle depending on their sizes.
The distribution has been studied previously \cite{Araki2002,Fujiyama20102},   especially with the expectation that it shows a self-similar structure. Fujiyama $et$ $al.$ found that the distribution obeys a self-similar distribution and corresponds to the fluctuation of the vortex line density \cite{Fujiyama20102}. In this study, we collect data on the lengths of rings and investigated the distribution and dependence on the anisotropy of the tangles.

\begin{figure}[h]
\includegraphics[scale=0.25]{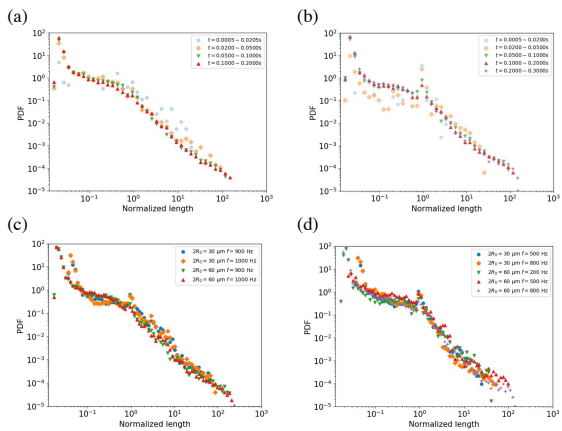}
\caption{\label{fig:vld}The vortex number density distribution. (a) and (c):``parallel injection'' (b) and (d):``spherical injection'' (a)-(b) are the time-development of the vortex number distribution averaged in the time interval in these legends. (c)-(d) are the averaged vortex number distributions after $0.2\ \mathrm{s}$ under different conditions.}
\end{figure}
The time developments of the distribution for ``parallel injection" and ``spherical injection" are shown in Fig. \ref{fig:vld}(a) and (b). The lengths are normalized by $2 \pi R_0$. In earlier times, the distribution concentrates in unity, referring to the initial injection vortices. The distributions develop into both smaller and larger than the injected length, which seems to correspond to a cascade process caused by vortex splitting and an inverse cascade process caused by vortex fusion. 
After a long time, it converges to a statistically steady distribution. 

It seems that the distributions are saturated just before their vortex line lengths reach a roughly constant value. This indicates the following picture about the development of vortex tangles. Initially, the tangle has vortex rings at a unique abundance ratio with respect to their sizes before the vortex line length is saturated. Consequently, the tangle develops while maintaining the abundance ratio of the vortex rings. Finally, both the ratio and vortex line lengths of the tangle are saturated. 

Figures \ref{fig:vld}(c) and (d) show the vortex length distributions under different  conditions. They show power laws, though the slopes below and above their injection length $R_0$ are different.  
Surprisingly, their lines are common, although the injection sizes, densities, and generating methods of the tangles are completely different.   It is important to investigate the universal properties of localized turbulence. 

The tangles by ``parallel injection" show the power law in the range above the injection size. These slopes are $-1.67\sim -1.98$. However, in the range below the injection size, any power law disappears. Too small vortex rings can be ignored because they are closer to the spatial resolution. 

The tangles by ``spherical injection" do not clearly show any power law. The distributions in the range above the injection sizes are similar to an exponential distribution. These values in the range below the injection sizes also do not show the power law.

According to the fractal theory of classical turbulence \cite{Frisch}, large eddies split into smaller eddies self-similarly, which is generally called the Richardson cascade, and the distribution of the number $N(l)$ of vortices with size $l$ obeys $N\propto l^{-D_f}$. However, the distribution in this simulation shows not only a cascade process but also an inverse cascade process. The correlation with the fractal dimension is not confirmed.

\section{conclusion}
We numerically investigate the fractal dimension and the vortex length distribution of the localized tangles as a consequence of our previous study \cite{Nakagawa2020}. 
The fractal dimension increases with the vortex line density in the case of low-density tangles, whereas the dimension of the high-density tangle is saturated. 
The vortex length distribution of the tangles by ``parallel injection'' shows the power law though the distribution of the tangle by ``spherical injection'' does not.

The aim of this series of studies is to associate the internal structure of the localized vortex tangle with the self-similarity of the vortex emission \cite{Yano2019,Nakagawa2020}. We considered that the vortex length distribution enabled us to discuss it quantitatively. However, the self-similarity of the vortex emission is observed in the range of approximately $3\ \mathrm{\mu m}$ to $60\ \mathrm{\mu m}$ in our previous study \cite{Nakagawa2020}, and the self-similarity of the vortex length distribution is not observed in this range. 
The statistics of the internal structure do not necessarily correspond to that of the emitted vortex rings.

 Although this simulation is performed at $T=0\ \mathrm{K}$, we will comment on the effects of finite temperatures. Kivotides $et$ $al.$ calculated the fractal dimension of vortex tangles at $1.3\ \mathrm{K}$ and $1.9\ \mathrm{K}$ and found that it did not depend on the temperature. When the mutual friction works at finite temperatures, the fine structure of a vortex tangle is smoothed out. A fine structure smaller than the intervortex distance does not affect the fractal dimension of the vortex tangles. Therefore, the fractal dimension should depend less on the temperature when it is given by a function of the vortex line length density.

\begin{acknowledgments}
M.T. was acknowledged by JSPS KAKENHI (Grant No. JP20H01855). S.I. was supported by a Grant-in-Aid for JSPS Research Fellows  (Grant No. JP20J23131).
\end{acknowledgments}

\bibliographystyle{apsrev4-2}
\bibliography{ref.bib}

\end{document}